\documentclass{article}
\usepackage{amsmath}
\usepackage{amssymb}

\input{tcilatex}
\begin{document}

\begin{center}
\textbf{Simple recursions displaying interesting evolutions}

\bigskip

\textbf{Francesco Calogero}

Dipartimento di Fisica, Universit\`{a} di Roma "La Sapienza", Rome, Italy

Istituto Nazionale di Fisica Nucleare, Sezione di Roma 1, Rome, Italy

Istituto Nazionale di Alta Matematica, Gruppo Nazionale di Fisica
Matematica, Italy

francesco.calogero@uniroma1.it, francesco.calogero@infn.roma1.it

\bigskip
\end{center}

\section{Introduction}

In this paper firstly the \textit{very simple nonlinear recursion }is
introduced:%
\begin{equation}
y_{n+1}=\left( y_{n}-1\right) ^{2}~,~~~n=0,1,2,...;  \label{1}
\end{equation}%
and---on the basis of rather elementary considerations---the main \textit{%
qualitative} features of its evolution are analyzed, firstly in the context
of \textit{real} numbers (including the subcase of \textit{rational}
numbers) and then in the more general context of \textit{complex} numbers.
Then---via a \textit{simple but nonlinear} change of dependent
variables---the following, \textit{related}, class of recursions, is
introduced:%
\begin{equation}
x_{n+1}=\frac{f_{2}\left( x_{n}\right) ^{2}+f_{1}x_{n}+f_{0}}{g_{2}\left(
x_{n}\right) ^{2}+g_{1}x_{n}+g_{0}}~,~~~n=0,1,2,...;  \label{2}
\end{equation}%
it features the $6$ \textit{a priori arbitrary coefficients} $f_{j}$ and $%
g_{j}$ ($j=2,1,0$), which however turn out to be expressed---by \textit{%
explicit} formulas---in terms of only $3$ \textit{arbitrary parameters} $%
a,b,c$; implying that the $6$ \textit{a priori arbitrary coefficients} $%
f_{j} $ and $g_{j}$ must satisfy---as it were, \textit{a posteriori}---$3$ 
\textit{constraints}, which are also reported below. The recursion thus
obtained---featuring of course only the $3$ \textit{arbitrary parameters }$%
a,b,c$---is also reported below; as well as a rather detailed \textit{%
qualitative} description of the behavior of its solutions.

The recursion (\ref{1}) is discussed in \textbf{Section 2}, and the
recursion (\ref{2}) (in fact, mainly the subcase of it mentioned just above)
in \textbf{Section 3}. Terse concluding remarks, including a \textit{hint}
at why these findings might be of broader scientific interest than just to
elementary number players, are in \textbf{Section 4}.

\textbf{Notation}: above and hereafter $n$ \ is a \textit{non-negative
integer}, playing the role of \textit{independent variable}: $n=0,1,2,3,...$%
; the index $j$ takes the $3$ values $2,1,0$; $y_{n}$ and $x_{n}$ are the 
\textit{dependent variables}, and they may be \textit{real} numbers, or
simpler or more complicated entities, as explained below. Below the
"approximate equality symbol" $\cong $ is occasionally used to indicate that
an equation involving \textit{real} numbers has been replaced by an \textit{%
approximate rational} approximation to it (for instance $\sqrt{2}\cong
7/5=1.4$: we of course always employ the \textit{decimal} system...). $%
\blacksquare $

\bigskip

\section{The first recursion}

The \textit{nonlinear recursion} investigated firstly is defined by the
following \textit{very simple} rule (see (\ref{1})):%
\begin{equation}
y_{n+1}=\left( y_{n}-1\right) ^{2}~.  \label{Iter}
\end{equation}%
For any assignment of the \textit{initial value} $y_{0}$ , the subsequent
values $y_{n}$ for $n=1,2,3$,...are clearly \textit{all} well defined: how
shall they evolve as $n$ increases? Note that this recursion implies that,
if the \textit{initial} value $y_{0}$ is an \textit{integer} or a \textit{%
rational} number, \textit{all} the subsequent numbers $y_{n}$ with $%
n=1,2,3,...$, shall also be \textit{integer} or \textit{rational} numbers.
Likewise if $y_{0}$ were a \textit{complex number}, all the subsequent
numbers $y_{n}$ would be \textit{complex numbers}: indeed an analogous
phenomenon would of course also characterize the evolution determined by the
recursion (\ref{Iter}) even in \textit{more general} cases, for instance if
the symbol $y_{0}$ were a \textit{square matrix }of \textit{integer (}or of 
\textit{real}, or of \textit{complex})\textit{\ numbers}, \textit{all} the
subsequent quantities $y_{n\text{ }}$ yielded by the recursion (\ref{Iter})
would also have the \textit{same} nature as $y_{0}$.

\bigskip

\subsection{The case of \textit{real} numbers}

An elementary observation is that, in this case, whichever is the initial
value $y_{0}$, the subsequent numbers $y_{n}$ with $n=1,2,3,...$shall 
\textit{all} be \textit{positive numbers}. Hence the analysis can be in this
case be restricted to the case of \textit{positive real numbers.}

Next we note that clearly, if the \textit{initial} value is $y_{0}=0$, the
recursion (\ref{Iter}) yields an extremely simple \textit{periodic} outcome$%
:y_{m}=1$ if $m$ is a \textit{positive odd integer,} $y_{m}=0$\ if $m$ is a 
\textit{positive even integer}. And clearly an \textit{analogous} outcome
obtains if the \textit{initial} value is $y_{0}=1$: then $y_{m}=0$\ if $m$
is a \textit{positive odd integer,} $y_{m}=1$\ if $m$ is a \textit{positive
even integer}.

But if the \textit{initial value} is a \textit{real} number inside the
interval $0<y_{0}<1$ or in its neighborhood, then the outcome is less
trivial. Indeed, if $y_{0}=\varepsilon $ with $0<\varepsilon <1$, then $%
y_{1}=\left( 1-\varepsilon \right) ^{2},$ $y_{2}=\varepsilon ^{2}\left(
2-\varepsilon \right) ^{2};$ etc. . It is therefore plain that, if $%
\varepsilon $ is a \textit{very small} number \textit{very close} to $0$,
the numbers produced by the recursion (\ref{Iter}) shall, at \textit{every}
step, jump from one side to the other side of the interval \ $0<y<1$,
getting at every consecutive step closer to its borders, but never actually
reaching them; except, so to say, in the limit as $n\rightarrow \infty $
(but this is of course not a proper limiting value of the sequence $y_{n}$,
as the value of $y_{n}$ gets, at every step of this sequence, alternatively
closer to the number $0$ or to the number $1$). And obviously this is also
true if the \textit{initial} value $y_{0}$ is close to the right end of the
interval $0<y_{0}<1$. But how close must the initial value $y_{0}$ be to one
or the other end of the interval $0<y_{0}<1$ in order for this kind of
behavior to prevail?

Let us (as it were, in a \textit{very naive} mood) investigate firstly what
happens when we start from the \textit{very middle} of this interval, namely
if we start from the value $y_{0}=1/2$. We then get the following sequence: 
\begin{subequations}
\begin{eqnarray}
y_{1} &=&\left( 1/2\right) ^{2}~=1/4~=0.25~, \\
y_{2} &=&\left( 3/4\right) ^{2}~=9/16=0.5625~, \\
y_{3} &=&\left( 7/16\right) ^{2}=49/256=0.191406~, \\
y_{4} &=&\left( 207/256\right) ^{2}=0.6538238525390625~,...
\end{eqnarray}%
These \textit{rational numbers }seem to, alternatively, approach from above
and from below the $2$ ends of the interval from $0$ to $1$. So \textit{one}
might \textit{bet} that also for \textit{all initial} \textit{real numbers}
in the interval from $0$ to $1$ that behavior shall characterize the
sequence (\ref{Iter}). But \textit{he} would loose that \textit{bet (}maybe 
\textit{she} would not be so \textit{naive}?).

Indeed, let us search next for the \textit{equilibria} of the recursion (\ref%
{Iter}). Such values $y_{n}$ of the dependent variable $y_{n}$ must of
course satisfy the \textit{algebraic} equation 
\end{subequations}
\begin{subequations}
\begin{equation}
y=\left( y-1\right) ^{2}~;
\end{equation}%
they are therefore the $2$ roots of the \textit{second-degree} \textit{%
algebraic equation} 
\begin{equation}
\left( y\right) ^{2}-3y+1=0~;
\end{equation}%
hence they are the $2$ \textit{real} (\textit{positive nonrational}) \textit{%
numbers} 
\end{subequations}
\begin{subequations}
\label{Equi}
\begin{equation}
\bar{y}_{\pm }=\left( 3\pm \sqrt{5}\right) /2~~,  \label{Equia}
\end{equation}%
\begin{equation}
\bar{y}_{-}\cong 0.381966011250~,~~~\bar{y}_{+}\cong 2.618033988749~.
\label{Equib}
\end{equation}%
And \textit{one} of these $2$ numbers happens to fall \textit{inside} the
interval from $0$ to $1$, while the other falls \textit{outside} it.

So we must proceed with our investigation. Let us begin by
investigating---in the context of \textit{real numbers} in the interval from 
$0$ to $1$---whether the \textit{equilibrium point} $\bar{y}_{-}$ of the
recursion (\ref{Iter}) is a \textit{stable} or \textit{unstable} equilibrium
point. To this end we introduce the change of variables from the set of 
\textit{real numbers} $y_{n}$ to the set of \textit{real numbers }$\eta _{n}$%
, 
\end{subequations}
\begin{subequations}
\begin{equation}
y_{n}=\bar{y}_{-}+\varepsilon \eta _{n}~;
\end{equation}%
here and hereafter we assume the parameter $\varepsilon $ to be \textit{%
infinitesimally small} and \textit{positive, }$\varepsilon >0$; this implies
that the \textit{real numbers} $\eta _{n}$ shall evolve according to the
following \textit{linear} recursion%
\begin{equation}
\eta _{n+1}=2\left( \bar{y}_{-}-1\right) \eta _{n}~,
\end{equation}%
which implies%
\begin{equation}
\eta _{n}=\eta _{0}\left[ 2\left( \bar{y}_{-}-1\right) \right] ^{n}\cong
\eta _{0}\left( -1.23606797748\right) ^{n}~,
\end{equation}%
(with $\eta _{0}$ an \textit{arbitrary real nonvanishing number}); and this
clearly entails that $\bar{y}_{-}$ is an \textit{unstable}---\textit{%
repulsive}---\textit{equilibrium point}, i. e. such that any sequence of 
\textit{real numbers} yielded by the recursion (\ref{Iter}) which starts 
\textit{sufficiently close} to the \textit{equilibrium point} $\bar{y}_{-}$
shall get \textit{away} from it, by becoming at every consecutive step a bit
farther from it (first to one side, next time to the other, and so on); and
so on for a while, until the values of $y_{n}$ get \textit{sufficiently close%
} to the boundary values of the interval from $0$ to $1$, for the regime
described above taking over, characterized by the fact that the values of $%
y_{n}$ get closer and closer to the boundaries of the interval from $0$ to $%
1 $, but always jumping from one side of that interval to the other at every
step.

We have thus completely described the peculiar behavior of the solutions of
the recursion (\ref{Iter}) in the \textit{real numbers} context, provided $%
0\leq y_{0}\leq 1$; and clearly that behavior would be analogous in the more
restricted context of \textit{rational numbers}, with the \textit{%
simplification} due to the fact that, \textit{in that context}, the \textit{%
unstable equilibrium} value $\bar{y}_{-}$ inside that interval does \textit{%
not} exist. Therefore, \textit{in that context}, the \textit{bet} mentioned
above would be \textit{won} rather than \textit{lost}! Related \textit{%
question}: would an Artificial Intelligence (AI) be able to identify \textit{%
this difference} among the sequences yielded by the recursion (\ref{Iter})
in the $2$ contexts of \textit{real} and \textit{rational} numbers? \textbf{%
Empirical finding}: it depends on \textit{how} you pose the question to
it/her/him; and also on how \textit{advanced} is the AI to whom you have
access.

To complete our analysis of the behavior of the solutions of the recursion (%
\ref{Iter}) in the \textit{real numbers} context, there remains to consider
the possibility that the initial value $y_{0}$ be larger than unity: $%
y_{0}>1 $. Then clearly, if $1<y_{0}<2,$ the next value $y_{1}$ ends up 
\textit{inside} the interval $0<y<1$, and we already know what happens next.
(And we also trivially know what happens if $y_{0}=2$). But we also know
that, a little beyond the value $y=2$, there is the \textit{second}
equilibrium point $\bar{y}_{+}$, see (\ref{Equi}). Is this a \textit{stable}
or an \textit{unstable} equilibrium point? To find out we make now the
following change of variables from the set of \textit{real numbers} $y_{n}$
to the set of \textit{real numbers }$\zeta _{n}$, by setting 
\end{subequations}
\begin{subequations}
\begin{equation}
y_{n}=\bar{y}_{+}+\varepsilon \zeta _{n}~,
\end{equation}%
where we again assume the parameter $\varepsilon $ to be \textit{%
infinitesimal }and \textit{positive}: this implies that now the \textit{real
numbers} $\zeta _{n}$ shall evolve according to the following \textit{linear}
recursion%
\begin{equation}
\zeta _{n+1}=2\left( \bar{y}_{+}-1\right) \zeta _{n}\cong \left(
3.6236067977498\right) \zeta _{n}~,
\end{equation}%
which implies%
\begin{equation}
\zeta _{n}=\zeta _{0}\left[ 2\left( \bar{y}_{+}-1\right) \right] ^{n}\cong
\zeta _{0}\left( 3.6236067977498\right) ^{n}~.
\end{equation}%
We therefore conclude that this is also an \textit{unstable} (i. e., \textit{%
repelling}) equilibrium point; hence if the \textit{initial value} is any 
\textit{real number} \textit{larger} than $\bar{y}_{+},$ $y_{0}>\bar{y}_{+}$
(implying $\zeta _{0}>0$), then the sequence of \textit{positive real numbers%
} $y_{n}$ shall indefinitely grow faster and faster, tending to $+\infty $
as the index $n$\ tends itself to $+\infty $; while if the \textit{initial
value }$y_{0}$ is any \textit{positive} \textit{real number} \textit{smaller}
than $\bar{y}_{+},$ $2<y_{0}<\bar{y}_{+}$ (implying $\zeta _{0}<0$), then
the sequence of numbers $y_{n}$ shall initially decrease in absolute value
until it approaches to value $2$. But when it is sufficiently \textit{close}
to $2$ (namely, just a bit \textit{smaller} than $1+\sqrt{2}$), at the next
step it shall end inside the interval from $0$ to $2$, and at the following
step it shall surely fall inside the interval $0<y<1$; and its next
evolution has been described above.

The qualitative description of the somewhat remarkable behavior of this
recursion is thus completely ascertained, in the context of the quantities $%
y_{n}$ being \textit{real numbers}.

Let us therefore \textit{summarize} our findings of the behavior of the
sequence of numbers yielded by the recursion (\ref{Iter}) in the context in
which the variables $y_{n}$ are \textit{real numbers }(the rest of this
paragraph is formulated so as to be understandable even to someone who has
not followed the details of the above analysis). If the \textit{initial
value }$y_{0}$ is an \textit{arbitrary negative real number}, $y_{0}<0$, the
number $y_{1}$ shall be a \textit{positive} \textit{number}, $y_{1}>0$,
hence this case is essentially included in the case in which the \textit{%
initial value} $y_{0}$ is\ itself \textit{positive}, $y_{0}>0$, see below;
but before considering that case, let us recall the \textit{special} case
when the \textit{initial value} $y_{0}$ \textit{vanishes}, $y_{0}=0$, hence
the recursion (\ref{Iter}) yields clearly the following \textit{periodic}
sequence of numbers: $y_{n}=1$ whenever the \textit{index} $n$ (which is by
definition a \textit{nonnegative integer number}) is an \textit{odd integer}%
, while $y_{n}=0$ when the \textit{index} $n$ is an \textit{even integer};
and let us also immediately note that the analogous \textit{periodic}
sequence obtains if the \textit{initial value} is instead $y_{0}=1$, except
that then $y_{n}=1$ when the \textit{index} $n$ is an \textit{even integer}, 
$y_{n}=0$ whenever the \textit{index} $n$ is an \textit{odd integer}. Next
we consider the cases in which $y_{0}$, hence \textit{all }the following
numbers\textit{\ }$y_{n}$ are \textit{positive numbers}, $y_{n}>0$. We begin
by recalling what happens if the initial value $y_{0}$ lies \textit{inside}
the interval from $0$ to $1$, $0<y_{0}<1$: then there exists a \textit{%
single unstable equilibrium point }of the sequence (\ref{Iter}),
characterized by the\textit{\ irrational real value }$\bar{y}_{-}$ (see (\ref%
{Equi})), such that if $y_{0}=\bar{y}_{-}$ then $y_{n}=\bar{y}_{-}$ for 
\textit{all} values of the index $n$ ($n=1,2,3,...$). Let us consider next
the case in which the \textit{initial value} is a \textit{real number}
positive but less then unity, $0<y_{0}<1$, but different from $\bar{y}_{-}$
; then the numbers $y_{n}$ shall jump at \textit{each} step of the sequence
from being \textit{smaller} to being \textit{larger} than $\bar{y}_{-}$, but
always remaining \textit{inside} the interval $0<y<1$: jumping at every step
from being closer to one and then the other end of that interval, but never
reaching those extremal values $0$ and $1$. The last case to be considered
in this summary is the case in which the initial value $y_{0}$ is a \textit{%
positive real} number larger than unity, $y_{0}>1$: there are then $3$
possibilities: if $y_{0}=\bar{y}_{+}$ (see (\ref{Equi})), then for\textit{\
all }values of the index $n$, $y_{n}=y_{0}=\bar{y}_{+}$\thinspace (\textit{%
equilibrium});\ if $1<y_{0}<\bar{y}_{+}$, then as the index $n$ increases,
the numbers $y_{n}$ shall eventually approach the value $1$ from above and
then enter the interval $0<y<1,$ and then behave as described above;
finally, if $y_{0}>\bar{y}_{+}$ , as the index $n$ increases, the
corresponding numbers $y_{n}$ shall become \textit{larger and larger},
increasing with jumps of \textit{increasing} lengths: being of course always 
\textit{positive} \textit{finite real numbers}, albeit characterized by the
property 
\end{subequations}
\begin{equation}
y_{n+2}-y_{n+1}>y_{n+1}-y_{n}>0~.
\end{equation}

This is the end of our \textit{complete} description of the \textit{%
qualitative} behavior of the sequence (\ref{Iter}) in the \textit{real
numbers} context.

Skeptical readers not convinced by the arguments presented up to here are of
course welcome to check by \textit{numerical computations}---easily
performable with the help of a PC---whether this description is indeed
correct...

\bigskip

\subsection{The case of \textit{complex} numbers}

The next interesting case is that in which the recurring numbers are instead 
\textit{complex numbers}, for which we adopt hereafter the notation%
\begin{equation}
z_{n}=y_{n}+\mathbf{i}u_{n}  \label{Complex}
\end{equation}%
where hereafter $\mathbf{i}$ is the \textit{imaginary} unit (so that $%
\mathbf{i}^{2}=-1$) and the $2$ numbers $y_{n}$ and $u_{n}$ are \textit{real
numbers }(of course the \textit{real numbers} $y_{n}$ should not be confused
with the \textit{real numbers} $y_{n}$ discussed above; except in the
special case in which $u_{0}=0$ hence $u_{n}=0$, when the \textit{complex}
numbers $z_{n}$ discussed hereafter reduce to the \textit{real} numbers
discussed above; a case we hereafter \textit{exclude} from consideration).

Then the recursion (\ref{Iter}) can be reformulated as follows: 
\begin{subequations}
\label{IterCompl}
\begin{equation}
z_{n+1}=\left( z_{n}-1\right) ^{2}~,
\end{equation}%
implying (via (\ref{Complex}))%
\begin{equation}
y_{n+1}=\left( y_{n}-1\right) ^{2}-\left( u_{n}\right) ^{2}~,
\end{equation}

\begin{equation}
u_{n+1}=2u_{n}\left( y_{n}-1\right) ~.
\end{equation}
The last formula written above implies that 
\end{subequations}
\begin{equation}
y_{n}=1+u_{n+1}/\left( 2u_{n}\right) ~,
\end{equation}%
and via this formula we then obtain, for the dependent variable $u_{n}$, the
recursion 
\begin{equation}
u_{n+2}=u_{n+1}\left\{ -2+\left( u_{n+1}\right) ^{2}/\left[ 2\left(
u_{n}\right) ^{2}\right] -2\left( u_{n}\right) ^{2}\right\} ~.
\end{equation}%
This formula suggests introducing the new dependent variable 
\begin{subequations}
\begin{equation}
w_{n}=\left( u_{n}\right) ^{2}~,
\end{equation}%
which shall then satisfy the recursion%
\begin{equation}
w_{n+2}=w_{n+1}\left[ w_{n+1}-4w_{n}\left( 1+w_{n}\right) \right]
^{2}/\left( 2w_{n}\right) ^{2}~.  \label{Iterw}
\end{equation}%
The \textit{initial-values} problem of this \textit{second-order recursion} (%
\ref{Iterw}) requires now the assignment of the $2$ \textit{initial data }$%
w_{0}$ and $w_{1}$.

Clearly if, for any \textit{arbitrary }(nonvanishing!) value of $w_{0}\neq 0$%
,~$w_{1}=0$, then $w_{n}=0$ also for $n=2,3,$...: this special \textit{%
pseudoequilibrium} solution of the recursion in the context of \textit{%
complex numbers }$z_{n}$ correspond to the\textit{\ periodic} solution in
the context of \textit{real} numbers $y_{n}$ discussed above, when $y_{n}$
jumps at every step from the value $0$ to the value $1$ and then from the
value $1$ to the value $0$ and so on and on indefinitely. We also get an 
\textit{analogous pseudoequilibrium }outcome ($w_{n}=0$ for $n=2,3,$...) if $%
w_{0}$ is again any \textit{arbitrary }(\textit{nonvanishing}!) \textit{real}
number and 
\end{subequations}
\begin{equation}
w_{1}=4w_{0}\left( 1+w_{0}\right) ~.
\end{equation}%
But in both cases, we would have just returned to the previously treated
case of \textit{real} numbers...

The next question to investigate in order to arrive at a \textit{qualitative}
understanding of the behavior of the sequence of \textit{complex numbers }$%
z_{n}$ produced by the recursion (\ref{IterCompl}) is to investigate whether
this recursion possesses a genuine \textit{equilibrium} solution, i. e. such
that%
\begin{equation}
y_{n}=\bar{y}~,~~~u_{n}=\bar{u}~,
\end{equation}%
where $\bar{y}$ is\ a\ \textit{real number} and $\bar{u}$ a \textit{%
nonvanishing real number}. It is then obvious that the \textit{third} of the 
$3$ eqs.\ (\ref{IterCompl}) implies $\bar{y}=3/2$, and then the \textit{%
second} of the $3$ eqs.\ (\ref{IterCompl}) implies $\left( \bar{u}\right)
^{2}=-5/4$, \ which is of course \textit{incompatible} with the requirement
that $\bar{u}$ be a \textit{real number}. We thus conclude that the
recursion (\ref{IterCompl}) does \textit{not have any equilibrium solution}
(of course in the context of \textit{complex numbers} which are \textit{not
real}).

Finally, it is rather easy to convince oneself that therefore the \textit{%
qualitative} evolution of the sequence produced by the recursion (\ref%
{IterCompl}) shall always be, in some sense, \textit{dominated} by the
behavior of its \textit{imaginary} component $u_{n}$; hence that \textit{%
eventually}---whatever the \textit{complex} initial datum $z_{0}=y_{0}+%
\mathbf{i}u_{0}$ (with $y_{0}$ and $u_{0}$ \textit{arbitrary real numbers},
except for the requirement $u_{0}\neq 0$)---then $\left\vert
z_{n}\right\vert \ $shall always become \textit{very large}, evolving 
\textit{asymptotically} (as the index $n$ becomes larger and larger: $%
n\rightarrow \infty $) \textit{approximately} as follows:%
\begin{equation}
z_{n+1}\simeq \left( z_{n}\right) ^{2}~,~~~y_{n+1}\simeq -\left(
y_{n}\right) ^{2}~,~\ ~u_{n+1}\simeq -2y_{n}u_{n}~,
\end{equation}%
implying that \ $\left\vert u_{n}\right\vert >>\left\vert y_{n}\right\vert $%
, that both $\left\vert y_{n}\right\vert $ and $\left\vert u_{n}\right\vert $
\textit{increase steadily} by steps each larger than the preceding one ($%
\left\vert y_{n+2}\right\vert -\left\vert y_{n+1}\right\vert >>\left\vert
y_{n+1}\right\vert -\left\vert y_{n}\right\vert >>0,~\left\vert
u_{n+2}\right\vert -\left\vert u_{n+1}\right\vert >>\left\vert
u_{n+1}\right\vert -\left\vert u_{n}\right\vert >0$), and that the numbers $%
y_{n}~$become eventually \textit{all negative} ($y_{n}<0$), while the signs
of the numbers $u_{n}\ $change at every step (so that $u_{n+1}u_{n}<0$).

This provides a rather detailed description of the \textit{qualitative}
behavior of the sequence of \textit{complex numbers} $z_{n}$ produced by the
recursion (\ref{IterCompl}) for any \textit{arbitrary} (complex!) \textit{%
initial value} $z_{0}$, when the index $n$ becomes \textit{sufficiently large%
}; or even for \textit{all} values of the index $n$ if the \textit{initial
value} $z_{0}$ is itself \textit{sufficiently large} (for instance the
requirement\ $\left\vert z_{0}\right\vert >1$ would be enough; of course
always provided $u_{0}$ does \textit{not }vanish!).

Let us finally emphasize that what we described here is the \textit{%
qualitative} behavior of the sequence of numbers produced by the recursions (%
\ref{Iter}) and (\ref{IterCompl}). Of course it is nowadays easy to
investigate \textit{numerically}---by computer---such very simple
recursions. But let us also emphasize that some of the \textit{qualitative}
findings reported above are \textit{rigorous mathematical statements}; while 
\textit{no} finding \textit{only} based on \textit{numerical experiments}
has \textit{any} \textit{rigorous mathematical validity} beyond the \textit{%
validity} of the \textit{numerical findings themselves}.

\bigskip

\section{The second recursion}

In the \textit{first} subsection of this \textbf{Section 3} it is shown
that, via a \textit{simple} yet \textit{nonlinear} change of dependent
variables, the recursion (\ref{2}) may be related to the recursion (\ref{1}%
), and the restrictions on the $3+3=6$ \textit{coefficients} $f_{j}$ and $%
g_{j}$ of the recursion (\ref{2}) that are implied by this development are
ascertained. Then, in the \textit{second} subsection of this \textbf{Section
3}, the \textit{qualitative} behavior of the solutions of the \textit{%
subclass} of the recursion (\ref{2}) identified in the \textit{first}
subsection is outlined, and an interesting \textit{subcase} is also treated.

\bigskip

\subsection{A simple nonlinear change of dependent variables}

In this subsection the \textit{simple} yet \textit{nonlinear} change of
dependent variables is introduced, from the variables $y_{n}$ satisfying the
recursion (\ref{1}) to the variable $x_{n}$ satisfying the recursion (\ref{2}%
): 
\begin{subequations}
\label{ChangeVaryx}
\begin{equation}
y_{n}=\left( ax_{n}+b\right) /\left( x_{n}+c\right) ~.  \label{Changea}
\end{equation}%
It involves the $3$ \textit{a priori arbitrary parameters }$a,~b,~c$.

This relation can be immediately inverted:%
\begin{equation}
x_{n}=-\left( cy_{n}-b\right) /\left( y_{n}-a\right) ~.  \label{Changeb}
\end{equation}

And by inserting these relations in the recursion (\ref{1}) satisfied by the
dependent variable $y_{n}$ there obtains the recursion (\ref{2}) satisfied
by the new dependent variable $x_{n}$, with the following definitions of the 
$6$ \textit{coefficients} $f_{j}$ and $g_{j}$: 
\end{subequations}
\begin{subequations}
\label{ffgg}
\begin{equation}
f_{2}=(a-1)^{2}c-b~,~~f_{1}=2c\left[ (a-2)b+\left( 1-a\right) c)\right] ,~\
f_{0}=c(b^{2}-3bc+c^{2})~,  \label{fff}
\end{equation}%
\begin{equation}
g_{2}=-(a^{2}-3a+1)~,~~g_{1}=-2\left[ (a-1)b+c-2ac\right]
~,~~g_{0}=-b^{2}+2bc+(a-1)c^{2}~.  \label{ggg}
\end{equation}

These relations may easily be inverted, for instance expressing the $3$ 
\textit{parameters} $a,~b,~c$ in terms of the $3$ \textit{coefficients} $%
f_{2},~g_{1},~g_{2},$ as follows: 
\end{subequations}
\begin{subequations}
\label{abcG}
\begin{eqnarray}
a &=&(3-G)/2~,  \label{a} \\
b &=&-\left[ 2f_{2}-2g_{1}+8f_{2}g_{2}+g_{1}g_{2}-\left(
2f_{2}+g_{1}g_{2}\right) G\right] /G_{2}~,  \label{b} \\
c &=&\left[ 2f_{2}+3g_{1}-4f_{2}g_{2}+\left( 2f_{2}+g_{1}\right) G\right]
/G_{2}~;  \label{c} \\
G &=&\sqrt{5-4g_{2}}~,~~~G_{2}=4g_{2}\left( 1+g_{2}\right) ~.  \label{GG}
\end{eqnarray}

Clearly, because the $6$ \textit{coefficients} $f_{j}$ and $g_{j}$ are
expressed in terms of \textit{only} $3$ \textit{parameters}, they cannot be
assigned \textit{all} $6$ \textit{arbitrarily}; they are indeed \textit{%
constrained} by the following $3$ relations, which express the $3$ \textit{%
parameters} $f_{0},~f_{1},~g_{0}$ in terms of $f_{2},~g_{1},~g_{2}$ (see
also (\ref{GG})): 
\end{subequations}
\begin{subequations}
\label{fffggg}
\begin{eqnarray}
f_{0} &=&F_{1}\left[ \left( F_{1}\right) ^{2}+3F_{1}F_{2}+\left(
F_{2}\right) ^{2}\right] /\left( G_{2}\right) ^{3}~,  \label{f0} \\
f_{1} &=&F_{1}\left[ \left( 5-G\right) F_{1}-1-GF_{2}\right] F_{2}/\left(
G_{2}\right) ^{2}~,  \label{f1} \\
g_{0} &=&\left[ \left( F_{1}\right) ^{2}\left( 1-G\right)
/2+2F_{1}F_{2}+\left( F_{2}\right) ^{2}\right] /\left( G_{2}\right) ^{2}~,
\label{g0} \\
F_{1} &=&2f_{2}\left( 1-2g_{2}\right) +3g_{1}+\left( 2f_{2}+g_{1}\right) G~,
\label{F1} \\
F_{2} &=&\left[ -2f_{2}\left( 1+4g_{2}\right) +g_{1}\left( 2-g_{2}\right)
+\left( 2f_{2}-g_{1}g_{2}\right) G\right] ~.  \label{F2}
\end{eqnarray}%
~,

\bigskip

\subsection{Qualitative behavior of the solutions of the (restricted)
recursion (\protect\ref{2})}

In this subsection we tersely describe the \textit{qualitative} behavior of
the solutions of the \textit{subclass} of the recursion (\ref{2}) identified
in the \textit{previous} subsection. This analysis is greatly facilitated by
the relations (\ref{ChangeVaryx}) (if need be, associated with the formulas (%
\ref{ffgg}) and (\ref{abcG})), since we already mastered a full
understanding of the \textit{qualitative} behavior of the solutions of the
very simple recursion (\ref{1}): see \textbf{Section 2}. For the sake of
simplicity, we hereafter operate in the context of \textit{real} numbers.

Let us first of all rewrite here the recursion (\ref{2}) in the simplified
form appropriate for the \textit{subclass} under present consideration. It
is then more convenient to use directly \textit{only} the $3$ \textit{a
priori arbitrary} \textit{parameters} $a,b,c$ (rather than the $6$ \textit{%
coefficients} $f_{j},g_{j}$ given in terms of these coefficients by the
formulas (\ref{ffgg})). It is thus easily seen that the \textit{new} version
of the recursion (\ref{2}) reads as follows: 
\end{subequations}
\begin{equation}
x_{n+1}=\frac{\left[ b-\left( 1-a\right) ^{2}c\right] \left( x_{n}\right)
^{2}+2c\left[ \left( 2-a\right) b-\left( 1-a\right) c\right] x_{n}-\left(
b^{2}-3bc+c^{2}\right) c}{\left( 1-3a+a^{2}\right) \left( x_{n}\right) ^{2}-2%
\left[ \left( 1-a\right) b-\left( 1-2a\right) c\right] x_{n}+\left(
1-a\right) c^{2}-2bc+b^{2}}~.  \label{abc2}
\end{equation}

The simpler finding---which now may appear somewhat less than trivial---is
that this \textit{nonlinear} recursion (\ref{abc2}) feature $2$ \textit{%
periodic} solutions, both with period $2$, so that 
\begin{equation}
x_{n+2}=x_{n}~;
\end{equation}%
these solutions correspond of course to the $2$ trivial \textit{periodic}
solutions of the simple recursion (\ref{1}) described at the very beginning
of \textbf{Subsection 2.1}; they are therefore produced by the $2$ following
assignments of \textit{initial} data (as yielded by the relations (\ref%
{ChangeVaryx}) with $y_{0}=0$ respectively $y_{0}=1$):%
\begin{equation}
x_{0}=-b/a\equiv x_{-}~~~\text{respectively~~~}x_{0}=\left( c-b\right)
/\left( a-1\right) \equiv x_{+}~.  \label{x0periodic}
\end{equation}

The skeptical reader is \textit{welcome} to check!

Next, let us also ascertain what are the $2$ \textit{equilibrium solutions }$%
\bar{x}_{\pm }$ of the recursion (\ref{2}), corresponding to the $2$ \textit{%
equilibrium solutions }$\bar{y}_{\pm }$ of the recursion (\ref{abc2}). They
are of course immediately obtained from the formula (\ref{ChangeVaryx}) with
(\ref{Equi}), implying: 
\begin{equation}
\bar{x}_{\pm }=-\left\{ \left[ c\left( 3\pm \sqrt{5}\right) -b\right] /\left[
3\pm \sqrt{5}-a\right] \right\} /2~.  \label{xbar+-a}
\end{equation}%
And again the skeptical reader is \textit{welcome} to check that, by
replacing $x_{n+1}$ with $\bar{x}_{\pm }$ in the \textit{left-hand} side of
the formula (\ref{abc2}), and likewise $x_{n}$ with $\bar{x}_{\pm }$ in the 
\textit{right-hand} side of that same formula (\ref{abc2}), there obtains an 
\textit{identity}: implying of course that these $2$ values $\bar{x}_{\pm }$
are indeed \textit{equilibria} of the recursion (\ref{abc2}).

Another (somewhat less than obvious) result, which does however follow quite
easily from the findings reported in \textbf{Subsection 2.1}, is that 
\textit{all }solutions $x_{n}$ of the \textit{nonlinear} recursion (\ref%
{abc2}) which start \textit{inside} the interval---of length $\left\vert
\left( ac-b\right) /\left[ a\left( a-1\right) \right] \right\vert $---having
as its borders the $2$~values $x_{\pm }$ (see (\ref{x0periodic})), do remain 
\textit{inside} that interval for all values of the index $n$ and approach,
steadily, but at \textit{alternating} times, the $2$ borders of that
interval; unless they happen to coincide with one of the $2$ (\textit{%
unstable}) equilibria (\ref{xbar+-a}). Note moreover that, because for $a=0$
or $a=1$ that interval becomes \textit{infinitely\ long}, we may also
conclude that \textit{all} solutions of either one of the following $2$
recursions (featuring now \textit{only} the $2$ \textit{arbitrary parameters}
$b$ and $c$) 
\begin{subequations}
\label{SpeRec}
\begin{equation}
x_{n+1}=\frac{\left( b-c\right) \left( x_{n}\right) ^{2}+2c\left[ 2b-c\right]
x_{n}-\left( b^{2}-3bc+c^{2}\right) c}{\left( x_{n}\right) ^{2}-2\left(
b-c\right) x_{n}+\left( c-b\right) ^{2}}~,  \label{Reca=0}
\end{equation}%
\begin{equation}
x_{n+1}=\frac{b\left( x_{n}\right) ^{2}+2cbx_{n}-\left(
b^{2}-3bc+c^{2}\right) c}{-\left( x_{n}\right) ^{2}-2cx_{n}+b\left(
b-2c\right) }~,  \label{Reca=1}
\end{equation}%
shall \textit{diverge} as $n\rightarrow \infty $ by jumping alternatively to
values larger and larger in modulus but featuring at each step alternating
signs; unless of course they start from the following $2$ \textit{respective}
(\textit{unstable}) \textit{equilibria} of these $2$ recursions: 
\end{subequations}
\begin{subequations}
\label{Eqila=0,1}
\begin{equation}
\bar{x}_{\pm }=-\left\{ \left[ c\left( 3\pm \sqrt{5}\right) -b\right] /\left[
3\pm \sqrt{5}\right] \right\} /2~,  \label{Eqila=0}
\end{equation}%
\begin{equation}
\bar{x}_{\pm }=-\left\{ \left[ c\left( 3\pm \sqrt{5}\right) -b\right] /\left[
2\pm \sqrt{5}\right] \right\} /2~.  \label{Eqila=1}
\end{equation}

Let us conclude by noting that the findings reported in this \textbf{%
Subsection 3.2} provide a complete, rather detailed, understanding of the 
\textit{qualitative} behavior of \textit{all} solutions of the recursion (%
\ref{abc2}) (and also of its $2$ \textit{subcases} (\ref{SpeRec})); by
simply translating via the simple relation (\ref{ChangeVaryx}) the \textit{%
qualitative} understanding---provided above, see \textbf{Subsection 2.1}%
---of the corresponding \textit{qualitative} behavior of the recursion (\ref%
{1}).

\bigskip

\section{Envoy}

This paper has been written in a \textit{colloquial mood}, as possibly
permitted to the \textit{very old} scientist I happen to be. It contains 
\textit{no references}, because the main technique employed is so \textit{%
elementary} that to report previous uses of analogous approaches would have
required \textit{many thousands} of \textit{references}, including also to
many previous papers by myself (alone or with co-authors); where the same
simple trick---just a \textit{change of dependent variables}---has been
employed, which has been crucial to get the main findings of this paper.
Indeed I disapprove the now prevailing fashion of presenting an \textit{%
excessive} number of references (even leading to the suspicion that they are
now becoming commodities subject to market laws).

While I hope that the \textit{main} findings reported in this
paper---namely: that the \textit{simple nonlinear recursion} (\ref{abc2})
(featuring $3$ \textit{a priori arbitrary parameters} $a,b,c$) is in fact 
\textit{easily manageable} (by techniques perhaps providing more insight
than mere numerical computations; perhaps even than those made more powerful
by the availability of computers also equipped with some AI)---will be of
some interest to \textit{applied} mathematicians, active in such fields as
economics, population dynamics, engineering,... . Indeed the fact that a
certain \textit{simple nonlinear recursion }is, rather easily, \textit{%
mathematically manageable}, entails $2$ possible kinds of \textit{applicative%
} interest: that recursion might \textit{describe} a phenomenon whose
evolution is of interest; or instead one might be interested to \textit{%
manufacture} a device or to \textit{create} a development which is described
by just that recursion.

\textbf{Final musing}: I will be curious, and perhaps able, to see---after
this paper gets somehow published---whether it will motivate younger
colleagues (perhaps in China!) to generate a flood of interesting
generalizations, obtained by applying \textit{analogous} approaches to
identify and investigate other, \textit{more complicated}, recursions than
the \textit{quite simple} ones treated above.

\bigskip

\section{Acknowledgement}

I like to thank my grandson Cor Langerak, who is an artist using AI, for the 
\textbf{Empirical finding} reported above.
\end{subequations}

\end{document}